\documentclass[print]{revtex4}
\usepackage{amsmath,amssymb}
\usepackage{graphicx}

\begin{document}

\title{\large\bf Quantum superchemistry in an output coupler of coherent matter waves}
\author{H. Jing$^{1,2}$ and J. Cheng$^{3}$}

\affiliation{$^1$Department of Physics, The University of Arizona,
1118 East 4th Street, Tucson, AZ 85721\\
 $^{2}$State Key Lab of Magnetic Resonance and Atomic
 and Molecular Physics,
 Wuhan Institute of Physics and Mathematics, CAS, Wuhan 430071, P.R.
 China\\
 $^3$Lab for Quantum Optics,
 Shanghai Institute of Optics and Fine Mechanics, Chinese Academy of Sciences,
 Shanghai 201800, P.R.China}

 \begin{abstract}

We investigate the quantum superchemistry or Bose-enhanced
atom-molecule conversions in a coherent output coupler of
matter waves, as a simple generalization of the two-color
photo-association. The stimulated effects of molecular
output step and atomic revivals are exhibited by steering
the rf output couplings. The quantum noise-induced
molecular damping occurs near a total conversion in a levitation trap.
This suggests a feasible two-trap scheme to make a stable
coherent molecular beam. \\


\end{abstract}

\baselineskip=16pt

\maketitle

\indent The realization of Bose-Einstein condensates (BEC) in
ultracold atomic gases has led to a profound revolution in
modern physics, from low-temperature physics to atom optics [1].
Recently there has been many interests in creating a molecular BEC
(MBEC) via a magnetic Feshbach resonance [2-3] or an optical
photo-association (PA) technique [4] within an atomic BEC. As an
atom-optics analog of the simplest nonlinear quantum optical
system (second-harmonic generation) [5], the quantum properties of
the hybrid BEC-MBEC have been extensively studied. For example,
Herbig ${\it et~ al.}$ [2] created a pure molecular quantum gas
spatially separated from its atomic partner by adding a magnetic
levitation field to an ordinary Feshbach resonance technique
within a cesium BEC. Winkler ${\it et~ al.}$ [4] experimentally
studied the coherent two-color PA process, a process first dubbed
by Heinzen ${\it et~ al.}$ as $superchemistry$ or Bose-enhanced
free-bound transition [6]. Also, Hope and Olsen [7] predicted the
onset of large molecular damping due to the quantum noises or
spontaneous processes in the quantum superchemistry where the
mean-field (MF) approach breaks down for certain parameters.

The coherent PA features a weak-light-induced resonant intra-mode
coupling and subsequent diatomic MBEC at low temperatures as the
microkelvin range, and the practical efficiency of molecules
generation is limited by radiative decay and particle collisions
[7]. The versatile technique of Feshbach resonance, in which the
spin of one of two colliding atoms flips in a magnetic field, is
widely exploited in an optical trap [2] to create a MBEC with
high efficiency [3], along with other important applications like
the generations of matter-wave bright solitons [8], which are made
possible by tuning the atomic interactions. By combining these
magneto-optic techniques, many theoretical schemes have been
proposed to achieve an efficient and stable conversion of atoms to
molecules in a trapped BEC [9], by minimizing the impact of negative
factors like the MF shift and the vibrational relaxations.

Recently, Zhang ${\it et~ al.}$ [10] proposed a Feshbach resonance
technique applied to a $travelling$ atomic beam instead of an optically
confined BEC to study atomic filamentation and pair correlations.
Then similarly, it is natural to extend to study the quantum
superchemistry in an atom-laser output coupler. In this article,
we consider the concrete example of the MIT rf output coupler which
first realized a pulsed atom laser to address the question of whether
the presence of an optical sheet tuned near the PA resonance immediately
below the magnetic trap [11] can lead to new and different quantum
dynamical effects in the output matter waves. The quantum superchemistry
of the out-coupled hybrid matter waves is studied by using the powerful
tool of positive-$P$ representation in quantum optics [7,12].
We identify the stimulated effects of the molecular output step and
atomic revivals by steering the rf output couplings. The coherent
conversion of atoms to molecules is robust, with the noise-induced
molecular damping occurring near the total conversion. If these effects are
typical, then the quantum superchemistry in an atom laser can be tunable
and potentially useful for, e.g., making a continuous molecule laser.
A practical realization of this scheme could be conceived by exploiting the
levitation-trap technique of Herbig ${\it et~al.}$ [2] to collect and then
separate the molecules from the output hybrid atomic-molecular beam.
 \noindent
\begin{figure}[ht]
 \setlength{\unitlength}{1.0mm}
             \centering
\includegraphics[width=0.6\columnwidth]{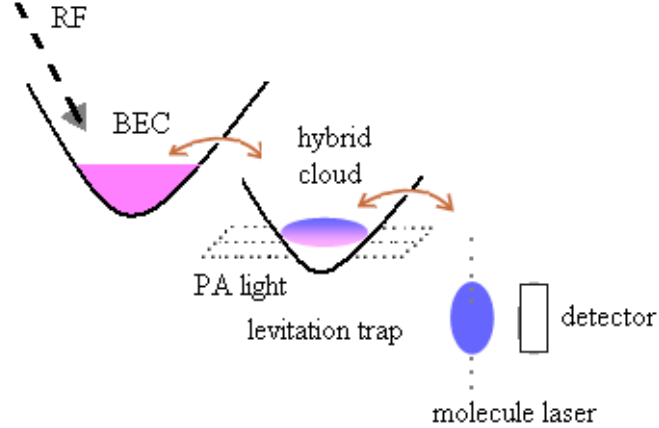}
\caption{(Color online) Schematic diagram of (non-Arrhenius) superchemistry
process in a matter wave coupler. The external RF field provides an effective
atomic $tunnelling$ between the two magnetic traps. This is followed by
molecule generation in the second (levitation) trap where the PA happens in
an initial $vacuum$. The hybrid BEC-MBEC collected in the levitation trap can
also be separated to form a coherent molecular beam [2, 11].} \label{3}
\end{figure}

As illustrated in Fig. 1, we consider for simplicity that a
large number of Bose-condensed atoms with two states is condensed
initially in the trapped state $|1\rangle $. State $|2\rangle $,
which is typically unconfined, is coupled to state $|1\rangle $ by
the rf field tuned near the $|1\rangle \rightarrow |2\rangle$
transition. The coherent PA process occurs in a levitation trap
which is initially in a $vacuum$. The resonantly enhanced atom-molecule coupling
then coherently converts [4] the output atoms into molecules in the
single-state $|3\rangle$. The Bose enhancement of the transition is
essential since the actual PA coupling is weak due to the low
Franck-Condon factors. Generally, to study the PA process one first
uses a single-mode approach (SMA) [13] to understand some main
features, especially for a one-dimensional BEC with small kinetic
energy or a coupling pulse of short duration. This approximation is
further evaluated by the recent work of Winkler ${\it et~ al.}$, who
found that their two-color PA experiment can be well described by a
SMA [4]. The spatial structure analysis can be an important
correction especially for large variations in density [7]. Here we
assume that both the atomic and molecular outputs form
controllable monochromatic beams and focus on the couplings of
different components. The boson annihilation operators for trapped
and output atoms, and subsequent molecules are denoted,
respectively, by $\hat{a}_1$, $\hat{a}_2$ and $\hat{b}$. Defining
the rf (Rabi) and PA strengths as $G'$ and $\gamma$, the effective
Hamiltonian in a rotating frame is then ($\hbar=1$)
\begin{equation}
\hat{H}=-\delta \hat{b}^{\dag}\hat{b}+G'(\hat{a}^{\dag}_1
\hat{a}_2+a^{\dag}_2 \hat{a}_1 )+\gamma(\hat{a}^{\dag 2}_2 \hat{b}
+\hat{b}^\dagger \hat{a}_2^2)+\lambda_a(\hat{a}^{\dag 2}_1
\hat{a}^2_1 + \hat{a}^{\dag 2}_2 \hat{a}^2_2)+\lambda_b
\hat{b}^{\dag 2} \hat{b}^2,
\end{equation}
where the molecular decaying into unimportant states can be
included by a non-Hermitian term in the optical detuning, e.g.,
$\delta \rightarrow \delta-\frac{1}{2}i\Gamma$ [13], with $\Gamma$
bing proportional to the total molecular decay rate. The
attractive intra-mode collisions are ignored in order to compare
with the trapped BEC case [7]. Our model is essentially
different from the well-known two-color PA [4, 6-7, 13] which, as
will be shown later, also can be included by our model under the
adiabatic approximation. Note that two different regimes may exist in our system,
the regime of atomic $tunnelling$, followed by molecule generation in the second
trap, and a propagating regime of separated matter waves (at the time
corresponding to the stable molecular step as described later). Our study focuses
on the first regime only and thus we ignored the kinetic-energy terms in the
Hamiltonian. For the long-distance propagating regime, Zhang $et~ al.$ already
predicted some novel effects for a travelling hybrid beam [8, 10, 18].

The full quantum effects are included by using the positive-$P$
representation of quantum optics [12]. The initial trapped state
$|1\rangle $ is taken as a coherent state and the other two
relevant states are initially in a vacuum. It is easy to represent them as $\delta$
functions but much difficult for an initial number state.
Using the standard correspondence relations between
$\{\hat{a}_1, \hat{a}^{\dag}_1, \hat{a}_2, \hat{a}^{\dag}_2,
\hat{b}, \hat{b}^{\dag}\}$ and $\{\alpha_1, \alpha^{\dag}_1,
\alpha_2, \alpha^{\dag}_2, \beta, \beta^{\dag}\}$,  we find the
following set of coupled $c$-number It\^o stochastic differential
equations $(\delta=0)$:
\begin{figure}[ht]
\includegraphics[width=0.6\columnwidth]{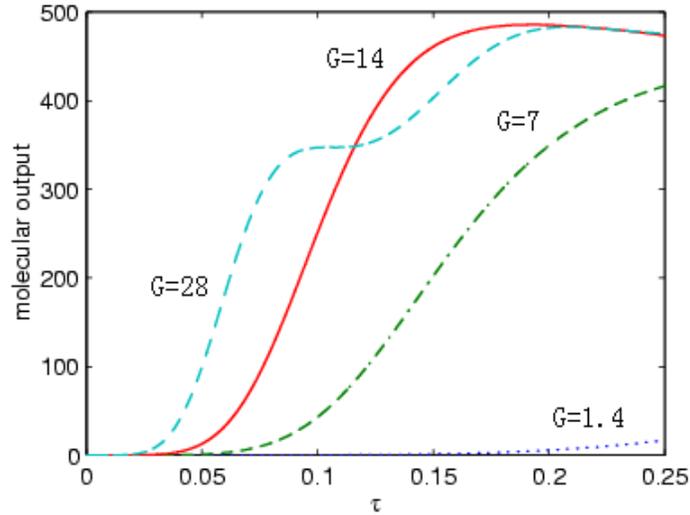}
\caption{(Color online) The molecular output numbers for different
rf couplings (scaled by the PA strength, i.e., $G=G'/\gamma$). The
scaled time $\tau$($=\gamma t)$ is dimensionless and the value
$0.145$ here corresponds to half of the maximum evolution time
$1\mu s$ used by Hope ${\it et~ al.}$ [7]. The parameters used
here are described in the text.} \label{2}
\end{figure}
\begin{eqnarray}
\frac{d\alpha_{1}}{dt} &=& -iG'\alpha_{2}-2i\lambda_a
\alpha_{1}^{+}\alpha_{1}^2
+\sqrt{-2i\lambda_{a}\alpha_1^{2}}\;\eta_{1}(t),
\nonumber\\
\frac{d\alpha_{1}^{+}}{dt} &=&
iG'\alpha_{2}^{+}+2i\alpha_{1}\lambda_a \alpha_{1}^{+\;2}
+\sqrt{2i\lambda_{a}\alpha_1^{+\;2}}\;\eta_{2}(t),\nonumber\\
\frac{d\alpha_{2}}{dt} &=& -iG'\alpha_{1}-2i\gamma
\alpha_{2}^{+}\beta -2i\lambda_a \alpha_{2}^2
\alpha_{2}^{+}+\sqrt{-2i\gamma\beta-2i\lambda_{a}\alpha_2^{2}}\;\eta_{3}(t),
\nonumber\\
\frac{d\alpha_{2}^{+}}{dt} &=& iG'\alpha_{1}^{+}+2i\gamma
\alpha_{2}\beta^{+} +2i\lambda_a \alpha_{2}^{+\;2}\alpha_{2}
+\sqrt{2i\gamma\beta^{+}+2i\lambda_{a}\alpha_2^{+\;2}}\;\eta_{4}(t),\nonumber\\
\frac{d\beta}{dt} &=&
-i\gamma\alpha_{2}^2-2i\lambda_{b}\beta^{+}\beta^{2}+\sqrt{-2i\lambda_{b}\beta^{2}}\;\eta_{5}(t),
\nonumber\\
\frac{d\beta^{+}}{dt} &=&
i\gamma\alpha_{2}^{+\;2}+2i\lambda_{b}\beta\beta^{+\;2}+\sqrt{2i\lambda_{b}\beta^{+\;2}}\;\eta_{6}(t),
\end{eqnarray}
where the $\eta_{i}$ are real Gaussian noise terms with the
correlations $\overline{\eta_{i}(t)}=0$ and
$\overline{\eta_{i}(t)\eta_{j}(t')}=\delta_{ij}\delta(t-t')$. The
$c$-number variables are not complex conjugate except in the mean
of a large number of stochastic trajectories. Eq.(2) can reduce to
the MF case in the noiseless limit. By solving it, any normally
ordered quantity can be calculated, including the quantum
statistics of the resultant fields which are difficult to be probed in
the experiment [4]. Here the quantities of interest are the
occupations in each of three states, especially the molecular output
number. The numerical stochastic integration is performed by averaging
selected moments of the fields over a sufficiently large sample of
trajectories. The initial atoms number is assumed as $1000$ and,
to compare with the trapped case [7], we set $\lambda_a=\lambda_b$
and use the mass and scattering length of $^{87}$Rb atoms.

We note that the scattering length of Rubidium molecules
and even the attractive (instead of repulsive) atom-molecule
interactions are now known experimentally. In fact, it is
straightforward to use some different collision values in our
numerical computation. (We have actually done this, and found that
the difference between the positive-$P$ and mean-field (MF)
results can be even greater in that case.) However, in order to
compare with the results of Hope and Olsen [7],  our present study
adopts their parameters, including the PA light strength and the
assumption of equal particle collisions. As shown in Fig.2, the
molecular output numbers generally increase with
respect to the scaled rf couplings; but for stronger rf coupling
($G=28$), we observe a molecular output step, i.e., the molecular
number increases rapidly at first but then much slowly. This has
a simple explanation. The rf coupling can induce a perfect atomic
oscillation without the PA [14] which, however, is deformed as an
atomic revival here. For a wide regime of rf couplings there is a
period of stable molecular output before reaching a total conversion,
with a similar time scale as in the trapped BEC case [7].

The new feature of the present scheme, as confronted in earlier
analysis, is that the coherent PA process starts not directly in
a trapped BEC, but in an output atomic field that is initially
in a $vacuum$. We note that the formally similar SMA model studied
by Calsamiglia ${\it et~ al.}$ [13] on two-color PA within a
Fock-state atomic BEC contains a linear molecular coupling and
their three-mode Hamiltonian was reduced by eliminating the
intermediate molecular mode under large evolution frequency
approximation. We focus here, however, on the role of $atomic$
tunnelling on the molecular generations in which the intermediate
atomic mode cannot be adiabatically eliminated [13].
 \noindent
\begin{figure}[ht]
\includegraphics[width=0.6\columnwidth]{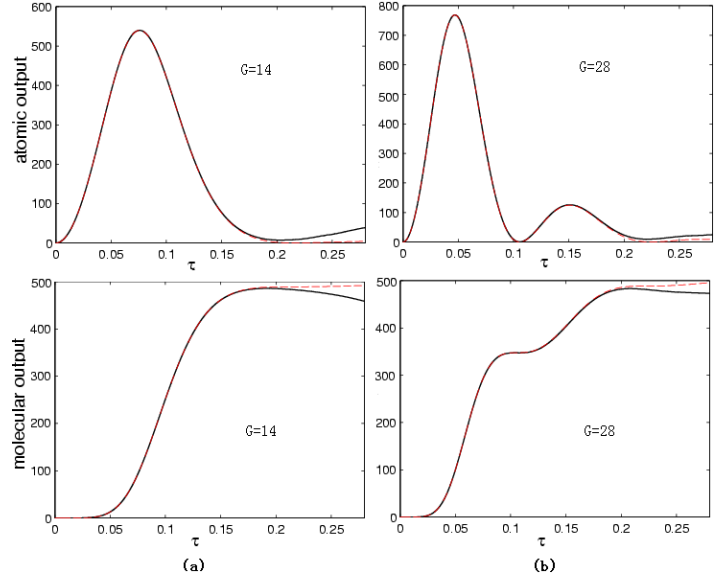}
\caption{(Color online)  The atomic and molecular output numbers
for different rf couplings. The black lines show the fully quantum
results while the red dotted lines are that of MF approach. For
lower rf coupling (a): the atomic revival is shown only by
considering the quantum noise [7]; but for stronger rf coupling
(b): more atomic revivals can appear in the output hybrid beams.
The first revival is exhibited even by MF approach. Around the
minimum atomic number, there is also an interesting output step
effect for the molecules.} \label{3}
\end{figure}

To compare the fully quantum and MF results, we plot the Fig.3 to
show the atomic and molecular output numbers for different rf
couplings. The MF results are simply from the noiseless limit of
Eq.(2). We see that, for weaker rf coupling, the molecular
generation is very similar as in the trapped BEC. But for stronger
rf coupling, an effect of molecular output step is observed around
the minimum atomic number, with an atomic revival instead of the
perfect oscillation in two-state stimulated transition [14]. This
provides clear evidence of a long-range coherence effect. In
contrast, the PA process in a thermal atomic beam would not
produce similar step or revival, because the phases associated
with the individual atom-molecule conversion are random in that
case. This is also different from the long-time behaviors
in a trapped BEC. These results, qualitatively valid beyond the
zero-dimensional model, clearly show the roles of the linear rf
output coupling and the nonlinear PA interaction.

The fully quantum approach is effectively the same as the MF
approach before reaching full atom-molecule conversion [7].
But their differences can be largely amplified in a complex
three-dimensional analysis by taking into account of many
non-ideal factors with the long-time evolutions [7,15]. Thus it
might be of some interests for an atom-laser-based experiment,
instead of a tightly confined BEC, to observe the molecular damping
first predicted by Hope and Olsen [7] in quantum superchemistry.

Our model also can include the quasi-bond-bond molecular coupling
$\Omega (\hat{g}^{\dag}\hat{b}+\hat{b}^{\dag}\hat{g})$ by applying
a pair of short, overlapping laser pulses, with $\hat{g}$ denoting
the stable molecular mode, as a generalization of the familiar
Raman two-color PA process [7,13]. In fact, if we ignore the
collisions in the rarely populated intermediate molecular mode and
take $|\delta|$ as the largest frequency in the system,
we can set $\dot{\hat{b}}/\delta=0$ and obtain
$\hat{b}\simeq(\Omega\hat{g}+\gamma\hat{a}_2^2)/{\delta}$. This
amounts to adiabatic elimination of the excited-molecular mode and
yields an effective interacting Hamiltonian which still has the
form of Eq. (1) with the following substitution (neglecting the free
motions and the tunable atomic collisions in the trap)
$$
\hat{b}\rightarrow\hat{g},~~G'\rightarrow G',~~\gamma\rightarrow
\chi=\gamma\Omega/\delta,~~ \lambda_a\rightarrow
\lambda_a+\gamma^2/\delta,~~\lambda_b\rightarrow \lambda_g.
$$
Hence, for $\gamma=145$ KHz, $\Omega=10$ GHz and $\delta=1000
~\Omega$, we have a much smaller effective PA strength $\chi=145$
Hz [7]. Due to $G=\eta G'/\gamma$ and $\eta=\delta/\Omega$, a large
scaled rf coupling requires a higher detuning. The maximum
strength here is chosen as: $G'\simeq 4$
KHz$<$$4\sqrt{6}$ KHz, in order to avoid the output shutdown effect
for an atom laser due to the density fluctuations in the
beam and the losses to other states [16]. This, together
with the molecular damping, also sets an upper limit to the flux
of a molecule laser. The possible noise suppression
due to, e.g., the dimer formation will be probed elsewhere.

In summary, we first reported the results of quantum superchemistry or
Bose-enhanced quantum dynamics in an output coupler of coherent
matter waves. We have showed that a step in the growth of the
molecular output and the atomic revivals can appear by steering
the external rf fields, and that the coherent atom-molecule
conversion is robust before the noise-induced molecular damping
occurring near a total conversion. It will add an experimental
level of difficulty to observe both two types of atomic revivals
due to the stimulated or the spontaneous processes, but with the
current progress in coherent coupling and PA of condensed atoms
[1,4,11], some possibilities are certainly open. If the effects of
molecular output step and rf-induced atomic revivals are typical,
then the superchemistry process in an atom laser is tunable and
very useful for achieving a stable output of molecules. The
technique of Herbig ${\it et~ al.}$ [2] and even a
collision-shift-compensated PA light with frequency chirp [9] also
could be applied to the hybrid beams in making, e.g., a continuous
molecule laser [11].

Our studies can be readily generalized to, e.g., a Raman output
coupler [17]. And the SMA is suitable for large couplings of
different components. In fact, the recent results of coherent PA
experiment of Winkler ${\it et~ al.}$ are well described by
the SMA [4]. We note that, for the long-distance propagating regime,
by using the MF nonlinear Schr\"odinger equation,
some novel effects were also predicted by Zhang $et~ al.$ for a travelling
hybrid beam with large attractive atom-molecule interactions, such as
the nonlocal behaviors or solitons of the output fields [8,10,18]. A
three-dimensional correction can be made similarly by using some extremely
complicated method and fully including many nonideal factors, such as the phase
diffusion of the BEC, the inelastic inter-mode scattering and even the
three-body losses at $higher$ densities (see, e.g., Refs.
[7,15,19-20]). Finally, the role of quantum statistics of the initial atoms in
our system will also be interesting and deserves further study.

\bigskip

\noindent H. J. thanks P. Meystre for his very useful discussions.
This work was supported by Natural Science Foundation of China
(10304020, 10404031), Wuhan Sunshine Plan and Shanghai Rising-Star
Program.

\end{document}